\documentstyle[12pt,titlepage]{article}
\input epsfig.sty

\font\grande=cmr9.5 scaled \magstep4
\font\medio=cmr9.5 scaled \magstep2
\outer\def\beginsection#1\par{\medbreak\bigskip
      \message{#1}\leftline{\bf#1}\nobreak\medskip
\vskip-\parskip
      \noindent}

\def\dis{\displaystyle}
\newcommand{\beq}{\begin{equation}}
\newcommand{\eeq}{\end{equation}}
\newcommand{\ud}{\mathrm{d}}
\newcommand{\beqn}{\begin{eqnarray}}
\newcommand{\eeqn}{\end{eqnarray}}
\newcommand{\nbeqn}{\begin{eqnarray*}}
\newcommand{\neeqn}{\end{eqnarray*}}
\newcommand{\bcen}{\begin{center}}
\newcommand{\ecen}{\end{center}}
\def\laq{\raise 0.4ex\hbox{$<$}\kern -0.8em\lower 0.62
ex\hbox{$\sim$}}
\newcommand{\benu}{\begin{enumerate}}
\newcommand{\eenu}{\end{enumerate}}
\newcommand{\bite}{\begin{itemize}}
\newcommand{\eite}{\end{itemize}}
\newcommand{\bdes}{\begin{description}}
\newcommand{\edes}{\end{description}}
\newcommand{\bdis}{\begin{displaymath}}
\newcommand{\edis}{\end{displaymath}}
\newcommand{\bary}{\begin{array}}
\newcommand{\eary}{\end{array}}

\newcommand{\bom}{\overline{\Omega}}
\newcommand{\bomax}{\overline{\Omega}^{\,{\rm max}}}
\newcommand{\bomth}{\overline{\Omega}^{\,{\rm th}}}

\def\gaq{\raise 0.4ex\hbox{$>$}\kern -0.7em\lower 0.62
ex\hbox{$\sim$}}

\begin{document}
\bibliographystyle {unsrt}

\titlepage

\vspace{15mm}
\begin{center}
{\grande Sensitivity of a VIRGO Pair to Relic GW Backgrounds }\\
\vspace{15mm}
D. Babusci $^a$ and M. Giovannini $^b$ 
\vspace{15mm}

{\sl $^a$ INFN- Laboratori Nazionali di Frascati, 1-00044 Frascati, 
Italy}\\

{\sl $^b$ Institute for Theoretical Physics, Lausanne University, }\\
{\sl BSP-Dorigny, CH-1015, Switzerland}
\end{center}

\vskip 2cm
\centerline{\medio  Abstract}

\noindent
The sensitivity of a pair of VIRGO interferometers to  
gravitational waves backgrounds (GW) of cosmological origin is analyzed 
for the cases of maximal and minimal overlap of the two detectors.
The improvements in the detectability prospects of scale-invariant 
and non-scale-invariant logarithmic energy spectra of relic GW  
are discussed.
\vspace{4cm}

\centerline{\bf Class. Quantum Grav. 17 (2000) 2621-2633}
\newpage
\section{Introduction and motivations}

Stochastic GW backgrounds \cite{gri} constitute 
a promising source for wideband interferometers whose operating 
window can be approximately located between few Hz and 10 kHz.
There are no compelling theoretical reasons why in such a frequency 
interval we should expect a negligible energy density stored in 
a stochastic background of primordial origin. Listening to 
phenomenology, we know that, unless $\Omega_{\rm GW}$ (the 
logarithmic energy spectrum of relic gravitons) is either flat or 
decreasing (as predicted, for instance, by some classes of 
inflationary models), the present constraints on stochastic 
gravitational waves backgrounds are quite mild. Listening to the 
theory we know that if $\Omega_{\rm GW}$ increases for frequencies 
larger than few mHz \cite{gio1}, then it is indeed possible to 
achieve a large signal in the operating window of interferometric 
detectors without conflicting neither with the fractional timing 
error of the millisecond pulsar's pulses \cite{pul} nor with the 
requirement that the total energy density of relic GW
should be smaller than the total amount of relativistic matter 
at nucleosynthesis \cite{ns}. These qualitative features can 
easily emerge in different models based on diverse physical 
frameworks including quintessential inflationary models \cite{quint},
 dimensional decoupling \cite{dim}, 
early violations of the dominant energy 
conditions \cite{dec} and superstring theories \cite{ven,sups}.

Recently the proposal of building in Europe an advanced interferometer 
of dimensions comparable with VIRGO \cite{vir} has been carefully 
scrutinized \cite{gia}. In view of this 
idea we would like to determine what would be the 
sensitivity of the correlation between two VIRGO-like detectors 
to a generic stochastic GW background. The answer to this 
question depends upon two experimental informations (i.e. the 
specific form of the noise power spectra and the relative location 
and orientation of the two detectors) and upon the theoretical 
form of the logarithmic energy spectrum. Concerning the last point, 
the specific frequency dependence of $\Omega_{\rm GW} (f)$ is 
extremely relevant. In fact $\Omega_{\rm GW}(f)$ directly enters 
in the expression of the signal-to-noise  ratio (SNR) \cite{ov1,ov2} 
and, therefore, logarithmic energy spectra with a different frequency 
dependence will lead, necessarily, to different SNR \cite{noi}.
 
Given a theoretical model, in order to compute reliably the 
sensitivity we have to specify the location of the two VIRGO detectors 
and the form of the noise power spectra. It is important 
to analyze how the relative distance between the two detectors of the 
pair can affect the sensitivity to the specific logarithmic energy 
spectrum of relic GW we ought to detect. In this respect, 
scale-invariant and non-scale-invariant logarithmic energy spectra 
lead to different sensitivity levels for the VIRGO pair already under 
the assumption that no improvement in the reduction of the thermal 
noises will take place prior to the construction of the second VIRGO 
detector. If a reduction in the contribution of the pendulum and 
pendulum's internal modes to the noise power spectra is achieved 
the improvements in the sensitivity will be even sharper \cite{us}.

Before starting our analysis we want to comment about the terminology. 
In our paper we will refer to GW. However, one should recall that 
we could also talk about relic graviton backgrounds. In fact 
the gravitational waves of cosmological origin are 
nothing but squeezed states of many gravitons produced 
from the vacuum fluctuations of the metric. In our investigations 
the quantum mechanical properties (i.e. correlations) of these 
states will not be essential but they can be, in principle, discussed along 
the lines of \cite{squ}.

\section{Maximal and minimal overlap} 

In the hypothesis that the GW background is isotropic and 
unpolarized the reduction of the sensitivity due to the relative 
distance and orientation of the two detectors depends upon the 
frequency $f$ and it is usually parameterized in terms of the 
(dimensionless) overlap reduction function \cite{ov2,alro}. 

By denoting (in spherical coordinates) with 
$\hat{\Omega}\,=\,(\cos{\phi}\,\sin{\theta},\,
\sin{\phi}\,\sin{\theta},\,\cos{\theta})$ 
a generic direction along which the given gravitational wave (GW) 
propagates, the overlap reduction function $\gamma(f)$ can be expressed, 
in terms of the pattern functions $F_i^A$ determining the response 
of the $i$-th detector ($i = 1,2$) to the $A = +,\times $ 
polarizations\footnote{We introduced the notation 
\bdis
\langle ... \rangle_{\hat{\Omega},\psi}\,=\,
\int_{S^2}\,\frac{\ud \hat{\Omega}}{4 \pi}\,
\int_0^{2 \pi}\,\frac{\ud \psi}{2 \pi}\,( ... )
\end{displaymath}
to denote the average over the propagation direction  
$(\theta,\phi)$ and the polarization angle $\psi$ of 
the GW.}:
\beqn
F^A (\hat{r},\hat{\Omega},\psi) &=& \mbox{Tr}\,\{ D (\hat{r})\,
\varepsilon^A (\hat{\Omega},\psi) \}
\label{pattern}\\
\gamma (f) &=& \frac{1}{F}\,\sum_A \,
\langle e^{i 2\pi f d\,\hat{\Omega} \cdot \hat{s}}\,
F_1^A (\hat{r}_1,\hat{\Omega},\psi) 
F_2^A (\hat{r}_2,\hat{\Omega},\psi) \rangle_{\hat{\Omega},\psi}\,=
\frac{\Gamma(f)}{F}.
\label{gamma}
\eeqn
Notice that $\Delta \vec{r} = \vec{r}_1 - \vec{r}_2 = d\,\hat{s}$ is 
the separation vector between the two detector sites and the 
summation over the index $A$ has to be taken over the physical 
polarizations (characterized by the polarization angle $\psi$)
\beqn
\varepsilon^+ (\hat{\Omega},\psi) &=& e^+ (\hat{\Omega})
\cos{2 \psi}\,-\,e^\times (\hat{\Omega}) \sin{2 \psi}\,, \nonumber \\
\varepsilon^\times (\hat{\Omega},\psi) &=& e^+ (\hat{\Omega})
\sin{2 \psi}\,+\,e^\times (\hat{\Omega}) \cos{2 \psi}\,,
\eeqn
of the incoming wave. The symmetric, trace-less tensor $D (\hat{r})$ 
appearing in Eq. (\ref{pattern}) depends on the geometry of the detector 
located in $\vec{r}$. A general analytical expression for the function 
$\gamma (f)$ can be written as  \cite{ov2,alro} 
\beq\label{gamten}
\gamma(f) = \rho_0 (\delta)\,D_1^{\,ij}\,D_{2\,ij}\,+\,\rho_1 (\delta)\,
D_1^{\,ij}\,D_{2\,i}^{\;k}\,s_j\,s_k\,+\,\rho_2(\delta)\,D_1^{\,ij}\,
D_2^{\,kl}\,s_i\,s_j\,s_k\,s_l
\eeq
where 
\beq\label{rhof}
\left[ 
\bary{c}
\rho_0(\delta)\\
\rho_1(\delta) \\
\rho_2(\delta)
\eary
\right]  = \frac{1}{F \delta^2}\,
\left[
\bary{rrr}
 2 \delta^2  & -4 \delta   & 2 \\
-4 \delta^2  &  16 \delta  & -20 \\
    \delta^2 & -10 \delta  & 35
\eary
\right]\
\left[
\bary{c}
j_0(\delta) \\
j_1(\delta) \\
j_2(\delta)
\eary
\right]\,,
\eeq
Notice that $j_k (\delta)$ are the standard spherical Bessel functions:
\bdis
j_0 (\delta) = \frac{\sin{\delta}}{\delta}\,, \qquad 
j_1 (\delta) = \frac{j_0 (\delta) - \cos{\delta}}{\delta}\,, \qquad
j_2 (\delta) = 3\,\frac{j_1 (\delta)}{\delta}\,-\,j_0 (\delta)\, ,
\edis
expressed as function of the dimensionless argument $\delta = 2 \pi f d$.

The normalization $F$ is given by
\beq
F = \sum_A\,\langle F_1^A (\hat{r}_1,\hat{\Omega},\psi) 
F_2^A (\hat{r}_2,\hat{\Omega},\psi) \rangle_{\hat{\Omega},\psi}
\,\mid_{1 \equiv 2}\,,
\label{F}
\eeq
where the notation $1 \equiv 2$ is a compact way to indicate that the 
detectors are coincident, coaligned and, if at least one of the two 
is an interferometer, the angle between its arms is equal to $\pi$/2 
(L-shaped geometry). In this situation, by definition, $\gamma (f) = 1$. 
When the detectors are shifted apart (so there is a phase shift between 
the signals in the two detectors), or rotated out of coalignment (so the 
detectors have different sensitivity to the same polarization)  
we will have that $|\gamma (f)| < 1$. 
The normalization factor $F$ is $2/5$ in the case 
of the VIRGO pair and in the case of any pair of wide band interferometers.

By varying the frequency $f$ the overlap reduction function is of order 
one until it reaches its first zero and, then, it starts oscillating around 
zero. The position of this first zero is roughly proportional to the 
inverse of the distance $d$ between the two detectors. The location 
of the corner station of the first interferometer will be assumed 
in Cascina (43.6 N , 10.5 E) where the VIRGO interferometer is 
presently under construction. The position of the second corner station is, 
at present, still under study. In our investigation we will suppose that 
the second corner station in different european sites. 

On a purely theoretical ground one ought to have a situation where the 
overlap reduction function is as close as possible to one for most of the 
frequencies in the operating window of the VIRGO pair. In other words we 
would like to push the first zero in the at higher frequencies
 because this would 
imply that the region of maximal overlap (i.e. $\gamma(f) \sim 1$ ) gets 
larger. Since an increase in the region of maximal overlap produces a 
decrease in the relative distance of the two interferometers it will not 
be possible to decrease the distance {\em ad libitum}. In fact, when we 
decrease the distance between the detectors, we  might introduce correlations 
between the local seismic and electromagnetic noises. We will assume that a 
distance of approximately $50$ km is sufficient to decorrelate these noises. 
At the moment we do not have indications against such an assumption.

The maximization of the overlap constitutes an important component as we can 
argue from the full expression of the signal-to-noise ratio. For the 
correlation of two interferometers, under the assumptions that the 
detector noises are Gaussian, much larger in amplitude than the 
gravitational strain and statistically independent on the strain itself, 
it can be shown \cite{ov1,ov2} that the signal-to-noise ratio in a frequency 
range $(f_{\rm m},f_{\rm M})$ is given, for an observation time $T$, by 
\footnote{ We follow the notations of Refs.\cite{noi,us}}
\beq
{\rm SNR}^2 \,=\,\frac{3 H_0^2}{5 \sqrt{2}\,\pi^2}\,\sqrt{T}\,
\left\{\,\int_{f_{\rm m}}^{f_{\rm M}}\,{\rm d} f\,
\frac{\gamma^2 (f)\,\Omega_{{\rm GW}}^2 (f)}{f^6\,S_n^{\,(1)} (f)\,
S_n^{\,(2)} (f)}\,\right\}^{1/2}\; ,
\label{SNR}
\eeq
where $H_0$ is the present value of the Hubble parameter. In Eq. (\ref{SNR}), 
the  performances achievable by the pair of detectors is controlled by the 
noise power spectra (NPS) $S_n^{\,(1,2)}$, whereas $\Omega_{\rm GW}(f)$ is 
the {\em theoretical} background signal defined through the logarithmic 
energy spectrum, normalized to the critical density $\rho_c$, and expressed at 
the present (conformal) time\footnote{In most of our equations we drop the 
dependence of spectral quantities upon the present time since all the 
quantities introduced in this paper are evaluated today.} $\eta_0$
\beq
\Omega_{{\rm GW}}(f,\eta_0)\,=\,\frac{1}{\rho_{c}}\,
\frac{{\rm d} \rho_{{\rm GW}}}{{\rm d} \ln{f}}\,=\, 
\overline{\Omega}(\eta_0)\,\omega(f,\eta_0)\,.
\label{Omegath}
\eeq
It is intuitively clear, from the combined analysis of the two previous 
expressions, that if $\gamma(f)$ reaches its first zero a low 
frequencies, the value of the integral will be smaller than in the case 
where $\gamma(f)$ reaches its first zero at larger frequencies. 

\section{Overlap versus noise power spectra}

\begin{figure}
\centerline{\epsfxsize = 7 cm  \epsffile{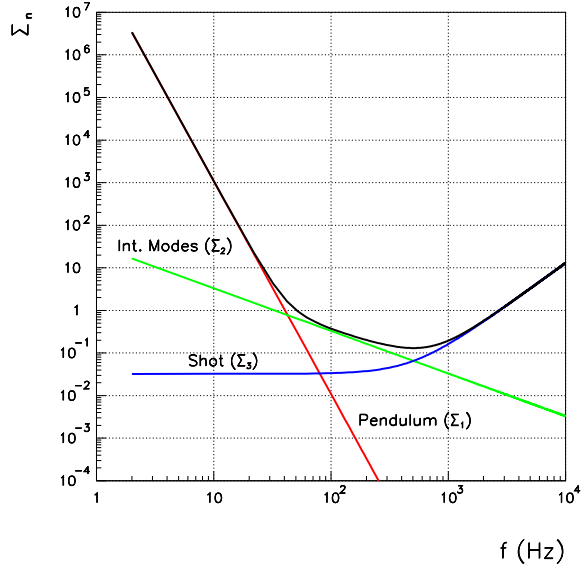}}
\vspace*{-1.5cm}
\caption[a]{The analytical fit of the rescaled noise power 
spectrum $\Sigma_n$ defined in Eq. (\ref{NPS}). 
With the full (thick) line the total NPS is reported.}
\label{noise}
\end{figure}

Given a specific  logarithmic energy spectrum of primordial GW the  
signal-to-noise ratio is determined by the interplay between the overlap 
reduction function and the noise power spectra of the detectors. The noise 
power spectrum of the VIRGO detector can be approximated by an analytical 
fit \cite{cuo}, namely 
\beq
\Sigma_n (f)\,=\,\frac{S_n (f)}{S_0}\,=\,
\left\{
\bary{lc}
\infty & \qquad \qquad f < f_b \\ [8pt]
\dis \Sigma_1\,\biggl(\frac{f_{{\rm a }}}{f}\biggr)^5\,+\,
\dis \Sigma_2\,\biggl(\frac{f_{{\rm a}}}{f}\biggr)\,+\,
\dis \Sigma_3\,\biggl[ 1 + \biggl(\frac{f}{f_{\rm a}}\biggr)^2\biggr],& 
\qquad \qquad f \ge f_b
\label{NPS}
\eary
\right.
\eeq
where 
\bdis
S_0\,=\,10^{-44}\,{\rm s}\;,\qquad f_a\,=\,500\,{\rm Hz}\;, 
\qquad f_b\,=\,2\,{\rm Hz}\;,\qquad
\bary{l}
\Sigma_1\,=\,3.46\,\times\,10^{-6} \\
\Sigma_2\,=\,6.60\,\times\,10^{-2} \\ 
\Sigma_3\,=\,3.24\,\times\,10^{-2}\,.
\eary
\edis
The noise power spectrum of the VIRGO detectors is reported in Fig. 
\ref{noise}.
In our parametrization, $\Sigma_1$ and $\Sigma_2$ control the 
contribution of the pendulum and pendulum's internal modes \cite{sau} 
to the thermal noise. $\Sigma_{3}$ controls instead the shot noise. 
For frequencies smaller than $f_b$ the noise power spectrum goes 
to infinity. The different values of $\Sigma_{1,2,3}$ together with 
$f_a$ and $f_b$ define a specific noise configuration of the detectors. 
If, in the future, one of the contributions to the noises will be reduced, 
the noise power spectra will change accordingly and the noise configuration 
might be different. Without entering into the details of the actual 
experimental method which could allow such a reduction we can parametrize 
possible changes in the noise power spectra through a noise ``reduction 
vector''
\beq
\vec{\rho} = ( \rho_1, \rho_2, \rho_3)
\eeq
where $\rho_i < 1$ defines the reduction in the corresponding coefficient 
$\Sigma_i$ entering Eq. (\ref{NPS}). Within the present noise 
configuration $\vec{\rho} = (1,1,1)$. 

By using into Eq. (\ref{SNR}) the expression of the theoretical 
spectrum given in Eq. (\ref{Omegath}) we have that the minimum 
normalization $\bom$ of a spectrum with functional variation 
$\omega(f)$, detectable by the VIRGO pair in an observation time 
$T$ with a given SNR is determined by 
\beq
h_{0}^2\,\overline{\Omega}\,\simeq\,\frac{4.0\,\times\,10^{-7}}{J}\;
\left(\,\frac{1\;{\rm yr}}{T}\,\right)^{1/2}\;{\rm SNR}^2\;.
\label{sens}
\eeq
In this equation the information of the specific $\omega(f)$ is 
encoded in the quantity $J$:
\beq
J \,= \biggl\{\,\int_{\nu_{\rm m}}^{\nu_{\rm M}}\,{\rm d} \nu\,
\frac{\gamma^2\,(f_0 \nu)\,\omega^2(f_0 \nu)}
{\nu^6\,\Sigma_n^{\,(1)} (f_0 \nu)\,
\Sigma_n^{\,(2)} (f_0 \nu)}\;\biggr\}^{\frac{1}{2}}
\label{Jint}
\eeq
where the integration variable is $\nu = f/f_0$ ($f_0$ is a 
generic frequency scale within the region 
$f_{\rm m}\,\le\,f\,\le\,f_{\rm M}$ ). We can assume  
$f_{\rm M}\,=\,10$ kHz, whereas 
the lower extreme $f_{\rm m}$ is put equal to the frequency $f_b$ 
entering Eq. (\ref{NPS}). The choice of $f_0$ is purely conventional 
and in view of our discussion we took $f_0= 100$ Hz.

Following the terminology of Sec. II the curve denoted by A in 
Fig. \ref{over} corresponds to the maximal overlap. The minimal overlap 
is represented by the curve C where the location of the second VIRGO 
detector coincides with the present location of the GEO \cite{geo} 
detector. For completeness we also illustrate a possible intermediate 
overlap (profile B) corresponding the situation where the two detectors 
are roughly $500$ km far apart. In principle, the effect of a 
maximization (or reduction) of the overlap between the two detectors 
of the VIRGO pair is not independent on the analytical form of 
the logarithmic energy spectrum $\omega(f)$. Indeed, as we will 
show in the following Sections a reduction in the overlap has 
a mild effect if the logarithmic energy spectrum is scale-invariant. 
However, if the logarithmic energy spectrum is non-scale-invariant 
(and it increases with frequency) then the effect can be sizable.

\section{Scale-Invariant Energy Spectra}

Suppose, as a warm-up, that $\omega(f) = 1$ so that the logarithmic 
energy spectrum is strictly scale-invariant. Then according to 
Eq. (\ref{sens}) the sensitivity of the VIRGO pair can be computed. 
\begin{figure}
\centerline{\epsfxsize = 7 cm  \epsffile{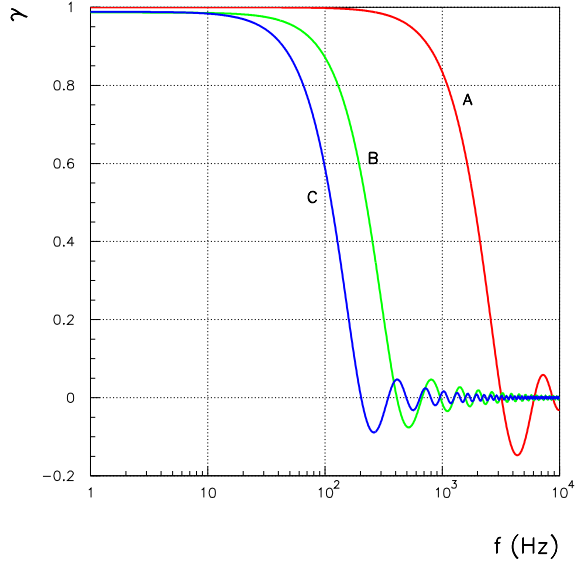}} 
\vspace*{-0.7cm}
\caption[a]{The overlap reduction function(s) for the 
correlation of the VIRGO detector presently under construction in 
Cascina (43.6 N, 10.5 E) with a coaligned interferometer whose (corner) 
station is located at: A) (43.2 N, 10.9 E), $d\,=\,58$ km (Italy); 
B) (43.6 N, 4.5 E), $d\,=\,482.7$ km (France); C) (52.3 N, 9.8 E), 
$d\,=\,958.2$ km (Germany).}
\label{over}
\end{figure}
Let us firstly assume that the two detectors have minimal overlap 
(i.e curve C in Fig. \ref{over}) and let us suppose that the VIRGO 
detectors will have the NPS given specifically by Eq. (\ref{NPS}). 
In other words the reduction vector will have its present value 
$\vec{\rho}=(1,1,1)$. In this case, from Eq. (\ref{Jint}) we can 
determine the sensitivity of the VIRGO pair. By using now into 
Eq. (\ref{sens}) the expression of the theoretical 
spectrum given in Eq. (\ref{Omegath}) we can determine the 
sensitivity of the VIRGO pair after one year of observation 
(i.e. $T = \pi \times 10^7$ s) and with SNR = 1 : 
\beq
h_0^2\,\bom \simeq 8.5 \times 10^{-8}\;.
\eeq
Suppose now to repeat the same estimate by assuming the 
maximal overlap. We want also to assume that the pendulum and 
pendulum's internal modes are suppressed by a factor of a hundred 
corresponding to a noise reduction vector $\vec{\rho}=(0.01,0.01,1)$.
Under this second set of assumptions we have that for $T = 1$ yr and 
SNR = 1 the sensitivity of the VIRGO pair becomes
\beq
h_0^2\,\bom \simeq 1.9 \times 10^{-9}\;.
\eeq
The two previous examples are quite extreme but a more complete
analysis of the interplay between maximization of the overlap and noise
reduction is illustrated in Tab. 1.
\begin{table}[!ht] 
\label{tab1}
\bcen
\caption{The minimum detectable $h_0^2\,\bom$ for the three different 
location of the second VIRGO detector (see Fig. \ref{over}) as a 
function of the reduction noise vector $\vec{\rho}$.}
\vspace*{0.2cm}
$$
\bary{|@{\quad} c @{\quad}||@{\quad} c @{\quad}|@{\quad} c @{\quad}|
@{\quad} c @{\quad}|}
\hline \rule{0ex}{3.5ex}
\vec{\rho} & A & B & C \\[5pt]
\hline \hline \rule{0ex}{3.5ex}
(1, 1, 1)       & 7.2\,\times\,10^{-8} & 7.6\,\times\,10^{-8} &
8.5\,\times\,10^{-8} \\[5pt]
(1, 1, 0.1)     & 6.9\,\times\,10^{-8} & 7.4\,\times\,10^{-8} &
8.3\,\times\,10^{-8} \\[5pt]
(0.1, 1, 1)     & 3.0\,\times\,10^{-8} & 3.1\,\times\,10^{-8} &
3.2\,\times\,10^{-8} \\[5pt]
(1, 0.1, 1)     & 2.5\,\times\,10^{-8} & 2.7\,\times\,10^{-8} &
3.4\,\times\,10^{-8} \\[5pt]
(1, 0.01, 1 )   & 1.8\,\times\,10^{-8} & 2.0\,\times\,10^{-8} &
2.5\,\times\,10^{-8} \\[5pt]
(0.01, 1, 1 )   & 1.2\,\times\,10^{-8} & 1.3\,\times\,10^{-8} &
1.3\,\times\,10^{-8} \\[5pt]
(0.1, 0.1, 1)   & 9.1\,\times\,10^{-9} & 9.6\,\times\,10^{-9} &
1.0\,\times\,10^{-8} \\[5pt]
(0.1, 0.01, 1 ) & 5.7\,\times\,10^{-9} & 6.0\,\times\,10^{-9} &
6.7\,\times\,10^{-9} \\[5pt]
(0.01, 0.1, 1 ) & 3.5\,\times\,10^{-9} & 3.6\,\times\,10^{-9} &
3.7\,\times\,10^{-9} \\[5pt]
(0.01, 0.01, 1) & 1.9\,\times\,10^{-9} & 1.9\,\times\,10^{-9} &
2.0\,\times\,10^{-9} \\[10pt]
\hline
\eary
$$
\ecen
\end{table}
In the first column of Tab. 1 we report the different values taken by the 
reduction vector $\vec{\rho}$ whereas in the second, third and 
fourth columns (from the left) we indicate the sensitivities 
achieved by a VIRGO pair under the assumption that the second 
(coaligned) VIRGO interferometer is located, respectively, in the three 
positions specified by the three profiles A, B and C of Fig. \ref{over}. 
The first corner station is always assumed, in our estimates, 
to be in Cascina.

In spite of the fact that the example of an 
{\em exactly} flat spectrum is academic and 
oversimplified \footnote{Needless to say that the 
analysis of the Sachs-Wolfe contribution to the detected amount 
of anisotropy implies, on a theoretical ground that, for 
$ f_{\rm m} < f < f_{\rm M} $, $h_0^2\,\Omega_{\rm GW} \leq 10^{-15}$ 
if a scale-invariant spectrum was originated during inflation.} 
we can draw, from our exercise, few interesting hints. We can see 
that if no noise reduction and  minimal overlap are simultaneously 
assumed the sensitivity is, comparatively, smaller than in the case 
where a selective reduction of the thermal noises and maximal overlap 
are postulated. If the reduction does not occur in the thermal components 
of the noise the improvement in the sensitivity is negligible, but 
still present. 

Our considerations were deduced only assuming a reduction in the 
thermal components of the noise. In principle, we should also consider 
the possible effect of a reduction in the shot noise.
If the reduction is selectively applied to the shot noise, the improvement 
in the sensitivity is negligible \cite{us}. 
This can be easily understood 
from the analysis of Fig. \ref{noise}: the shot noise starts being the 
dominant contribution to the NPS for $f \sim 1$ kHz, i.e. in a 
frequency region where the overlap begins to deteriorate (see 
Fig. \ref{over}). As far as the achievable sensitivity levels 
are concerned, we can notice that the three 
location are practically indistinguishable. This statement is even 
more accurate if the reduction in the thermal noise components gets 
larger.

For sake of completeness in Tab. 2 we report the minimum 
detectable $h_0^2\,\bom$ ($T$ = 1 yr and SNR = 1) for the correlation 
of VIRGO with the major interferometers presently under construction 
(LIGO, GEO, TAMA) \cite{int}. We also report the same quantity 
for the correlation of VIRGO with the existing resonant bars which are 
located close to the VIRGO site (AURIGA and NAUTILUS \cite{vcco}). 
These sensitivities are deduced from the calculation 
of ref. \cite{alro} in the case of interferometers, while are directly 
taken  from ref. \cite{vcco} for the resonant bars.
\begin{table}[!h] 
\label{tab2}
\bcen
\caption{The minimum detectable $h_0^2\,\bom$ for the correlation of 
VIRGO with other GW detectors in the case of flat spectrum ($T = 1$ yr, 
SNR = 1).}
\vspace*{0.2cm}
$$
\bary{|@{\quad} c @{\quad}||@{\quad} c @{\quad}|}
\hline \rule{0ex}{3.5ex}
{\rm VIRGO * LIGO-LA} & 1.6\,\times\,10^{-6} \\[5pt]
{\rm VIRGO * LIGO-WA} & 1.9\,\times\,10^{-6} \\[5pt]
{\rm VIRGO * GEO}       & 2.2\,\times\,10^{-6} \\[5pt]
{\rm VIRGO * TAMA}      & 3.6\,\times\,10^{-5} \\[5pt]
\hline \hline \rule{0ex}{3.5ex}
{\rm VIRGO * AURIGA}    & 1.6\,\times\,10^{-4} \\[5pt]
{\rm VIRGO * NAUTILUS}  & 2.8\,\times\,10^{-4} \\[10pt]
\hline
\eary
$$
\ecen
\end{table}

In ref. \cite{vcco} is also considered the possibility to correlate 
VIRGO with a (hyphotetical) spherical detector with a 3 m diameter 
made of Al 5056 ($M$ = 38 ton). The sensitivity turns out to be   
$h_0^2\,\bom \sim 2\,\times\,10^{-5}$ if the sphere is located in 
the AURIGA site, and it gets worse ( by a factor 2) if the sphere is 
moved to the NAUTILUS location. All these values have to be compared with 
the sensitivities reported in the first row of Tab. 1. Even in the  
less favorable case (correlation C of Tab. I) the sensitivity is greater (by
one order of magnitude) than the best achievable sensitivities reported 
in Tab 2.

\section{Non-scale-invariant spectra}
It is interesting to repeat a similar analysis in the case where 
$\omega (f)$, instead of being scale-invariant, increases with $f$.
In principle we would expect that in the latter case the impact of 
the maximization of the overlap will be more pronounced. The reason 
is that the contribution of $\omega(f)$ to the integrand of 
Eq. (\ref{SNR}) (or of Eq. (\ref{Jint})) increases at high frequencies 
if $\omega(f)$ increases. Therefore, if the first zero of $\gamma(f)$ 
falls just after 100 Hz (or possibly even before) the contribution 
of $\omega(f)$ will be erased more efficiently.

In order to show this behaviour let us analyze some specific examples 
among the ones mentioned in the introduction. For instance, in string 
cosmological models, the minimal pre-big-bang spectra have a two-branch 
form which can be expressed as \cite{ven,sups,us}
\beq
\omega (f)\,=\,
\left\{
\begin{array}{lc}
\dis z_s^{- 2 \beta}\,\left(\,\frac{f}{f_s}\,\right)^3\,\left[\,1\,+\,
z_s^{2 \beta - 3}\,-\,\frac12\,\ln{\frac{f}{f_s}}\,\right]^2 
& \dis \qquad \qquad f\,\le\,f_s\,=\,\frac{f_1}{z_s} \\ [15pt]
\dis \left[\,\biggl(\frac{f}{f_1}\biggr)^{3 - \beta}\,+\,
\biggl(\frac{f}{f_1}\biggr)^{\beta}\,\right]^2 
& \qquad \qquad f_s\,<\,f\,\le\,f_1
\eary
\right.
\label{minth}
\eeq
where 
\beq
\dis \beta\,=\,\frac{\ln\,(g_1/g_s)}{\ln\,z_s}\;,
\eeq
In this formula $z_s\,=\,f_1/f_s$ and $g_s$ are, respectively, the 
red-shift during, and the value of the coupling constant at the 
beginning of, the string phase \cite{ven,sups,us}. The maximal amplified 
frequency $f_1$ of the GW spectrum is
\beq
f_{1}(\eta_0)\,\simeq\,64.8\,\sqrt{g_1}\,\left(\,\frac{10^{3}}
{n_r}\,\right)^{1/12}\; {\rm GHz}\;,
\label{f1}
\eeq
where $n_{r}$ is the effective number of spin degrees of freedom in 
thermal equilibrium at the end of the stringy phase 
(of the order of 10$^2$ $\div$ 10$^3$, depending upon the 
specific string model), 
and $g_1\,=\,M_{s}/M_{\rm Pl}$ is the mismatch between the string 
($M_s$) and Planck ($M_{\rm Pl}$) masses. The value of $g_1$ 
corresponds to the dilaton coupling at the beginning of the radiation 
dominated epoch and it can be estimated to lie between $0.3$ and $0.03$
\cite{kap}.

In the case of increasing logarithmic energy spectra the evaluation of the 
sensitivity is a bit different from the case of purely flat spectra. If 
$\omega(f)$ grows in frequency the integrated spectrum can become, in 
principle large. We know, however, that the total amount of GW present 
at big-bang nucleosynthesis (BBN) cannot exceed the total amount of 
relativistic matter \cite{ns}. Otherwise the expansion rate of the Universe 
would increase too much and the observed light elements abundances could 
not be correctly reproduced. This implies that 
\beq
h^2_0\,\int_{f_{\rm ns}}^{f_{\rm max}}\,
\Omega_{\rm GW}(f,\eta_0)\;{\rm d}\ln{f}\,<\,
0.2\,h_0^2\,\Omega_{\gamma}(\eta_0)\,\simeq\,5\,\times\,10^{-6},
\label{ns}
\eeq
where $\Omega_{\gamma}(\eta_0)\,=\,2.6\,\times\,10^{-5}\,h_0^{-2}$
is the fraction of critical energy density stored in radiation at the 
present observation time $\eta_0$; $f_{\rm ns}\sim 10^{-10} $ Hz and 
$f_{\rm max}$ are, respectively, the nucleosynthesis frequency and 
the maximal frequency of the GW spectrum (for our present example 
$f_{\rm max} = f_1$). It is intuitively clear that if the spectrum is 
flat the BBN bound will be easily satisfied provided the theoretical 
amplitude of the spectrum, $\bomth$ is roughly less than $10^{-6}$. 
In the case 
of growing spectra the situation is more tricky. Let us denote with
\beq
h_0^2\,\bomax\,\simeq\,\frac{5\,\times\,10^{-6}}{{\cal I}}\;, 
\qquad {\cal I}\,= \,\int_{f_{\rm ns}}^{f_{\rm max}}\,
\omega(f)\;{\rm d}\ln{f},
\label{NSnorm}
\eeq
the maximal normalization of the spectrum compatible with the BBN bound. 
The sensitivity to a given $\omega(f)$ (which now increases with $f$) 
will be always given by Eq. (\ref{sens}). We should however exclude 
from the parameter space of our theoretical model those regions where 
$\bom > \bomax$. Thus, the regions of the parameter space of a given 
model for which $\bomax > \bom$ are, simultaneously, visible by the 
VIRGO pair and compatible with the BBN. In the absence of a specific 
theoretical normalization this would be the end of the story. However, 
in some models one can also estimate (with a given accuracy) the 
theoretical normalization of the spectrum. For instance, in the case 
of string cosmological models, the theoretical normalization of the 
spectrum can be expressed as  
\beq\label{theo}
\bomth \simeq 2.6\,g_1^2\,\left( \frac{10^{3}}{n_r}\right)^{1/3}\, 
\Omega_{\gamma}(\eta_0)\;. 
\eeq
The accuracy of this determination coincides with the accuracy in the 
determination of $g_1$ (the dilaton coupling at the beginning of the 
string phase) and of $n_r$ as defined in Eq. (\ref{f1}). 

If we want to compare the achievable sensitivity not only with the BBN 
bound but also with the theoretical normalization we will have to 
require that $\bomax > \bom $ and $\bomth > \bom$ are simultaneously 
satisfied. With these necessary specifications let us now analyze the 
impact of the maximization of the overlap in the case of 
non-scale-invariant spectra of string cosmological type. Our results 
for the ratios $\bomax/\bom$ and $\bomth/\bom$ are reported in 
Figs. \ref{ita}, \ref{ger}, and \ref{fra}. In Fig. \ref{ita} we illustrate 
the sensitivity to string cosmological spectra under the assumption of 
maximal overlap (profile A of Fig. \ref{over}). The region in the left 
plot of Fig. \ref{ita} traces the area of the parameter space where 
$\bomax > \bom$, whereas the right plot corresponds to the region of 
the parameter space for which $\bomth > \bom$. 
\begin{figure}[!hb]
\vspace*{-3.5cm}
\centerline{\epsfxsize = 21.5 cm  \epsffile{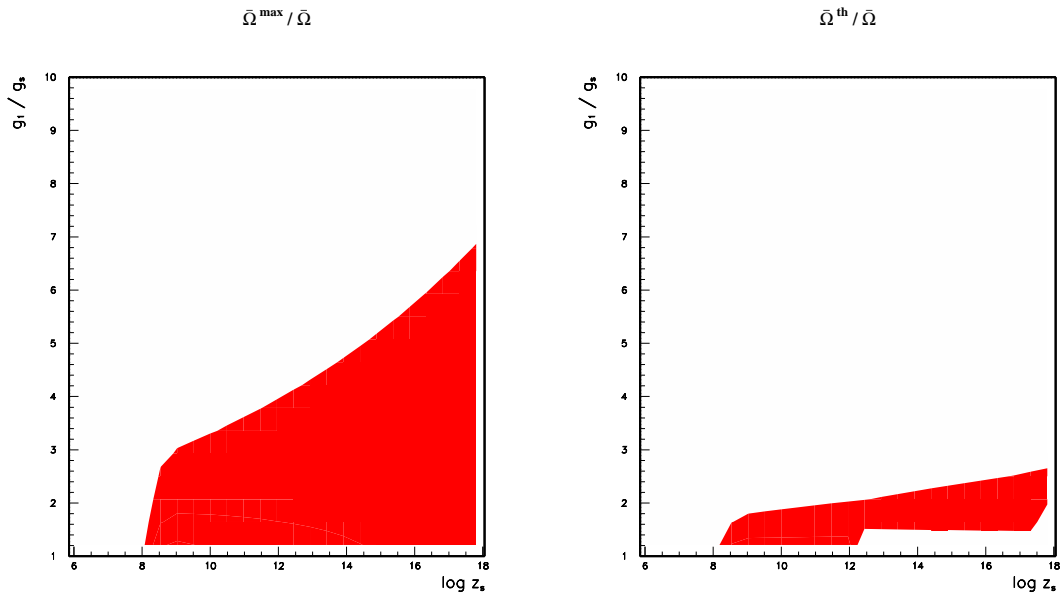}}
\vspace*{-18.0cm}
\caption[a]{ We report the regions 
for which  $\bomax/\bom > 1$ (left) and $\bomth/\bom > 1$ (right). 
The overlap is assumed to be maximal (profile A of 
Fig. \ref{over}). The fiducial set of parameters chosen for 
both plots $g_1 = 1/20$ and $n_r = 10^{3}$.}
\label{ita}
\end{figure}
The plots of Fig. \ref{ita} can be compared with the case where the overlap 
is not maximal. This is done in Figs. \ref{ger} and \ref{fra}.
\begin{figure}[!htp]
\vspace*{-3.5cm}
\centerline{\epsfxsize = 21.5 cm  \epsffile{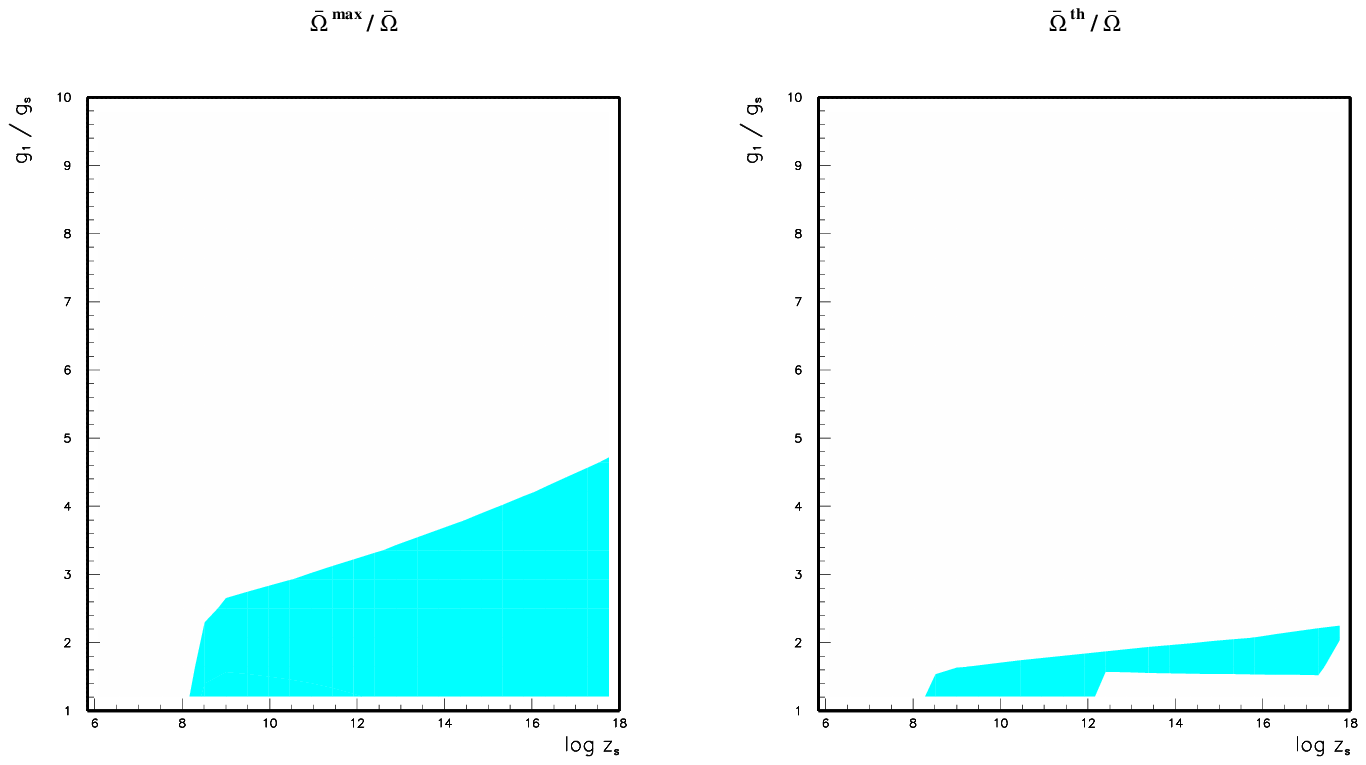}}
\vspace*{-18.0cm}
\caption[a]{We report the same quantities discussed in Fig. \ref{ita}, 
with the same fiducial choice  of $g_1$ and $n_r$ but under 
the assumption of minimal overlap, 
i.e. profile C in Fig. \ref{over}.}
\label{ger}
\end{figure}
\begin{figure}[!hbp]
\vspace*{-3.5cm}
\centerline{\epsfxsize = 21.5 cm  \epsffile{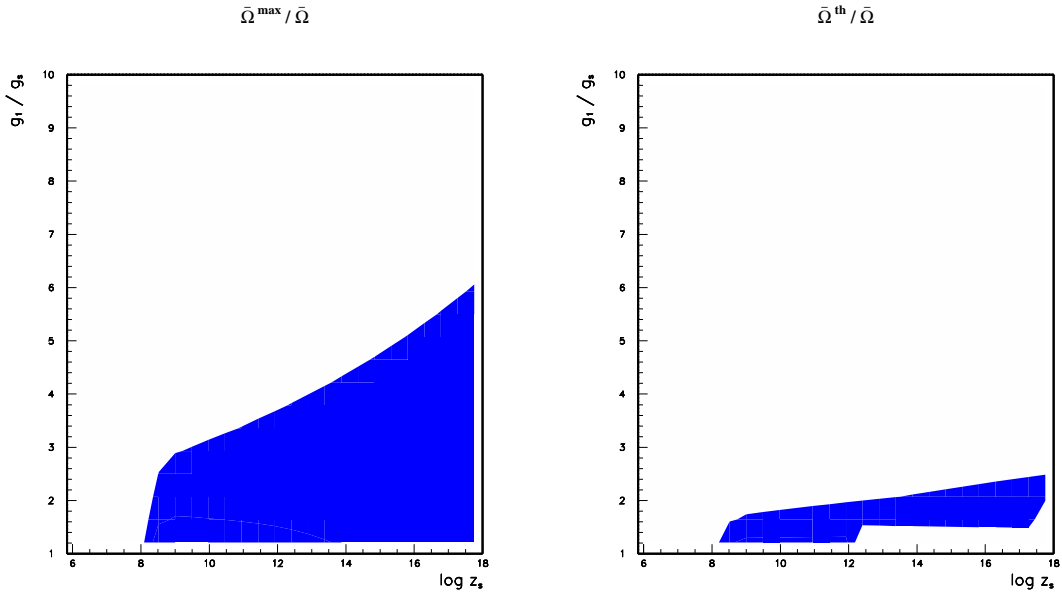}}
\vspace*{-18.0cm}
\caption[a]{We report the same visibility region discussed 
in Fig. \ref{ita} and \ref{ger} but in the case 
of intermediate overlap, i.e. profile B in Fig. \ref{over}.}
\label{fra}
\end{figure}
\begin{figure}[!hb]
\vspace*{-3.5cm}
\centerline{\epsfxsize = 21.5 cm  \epsffile{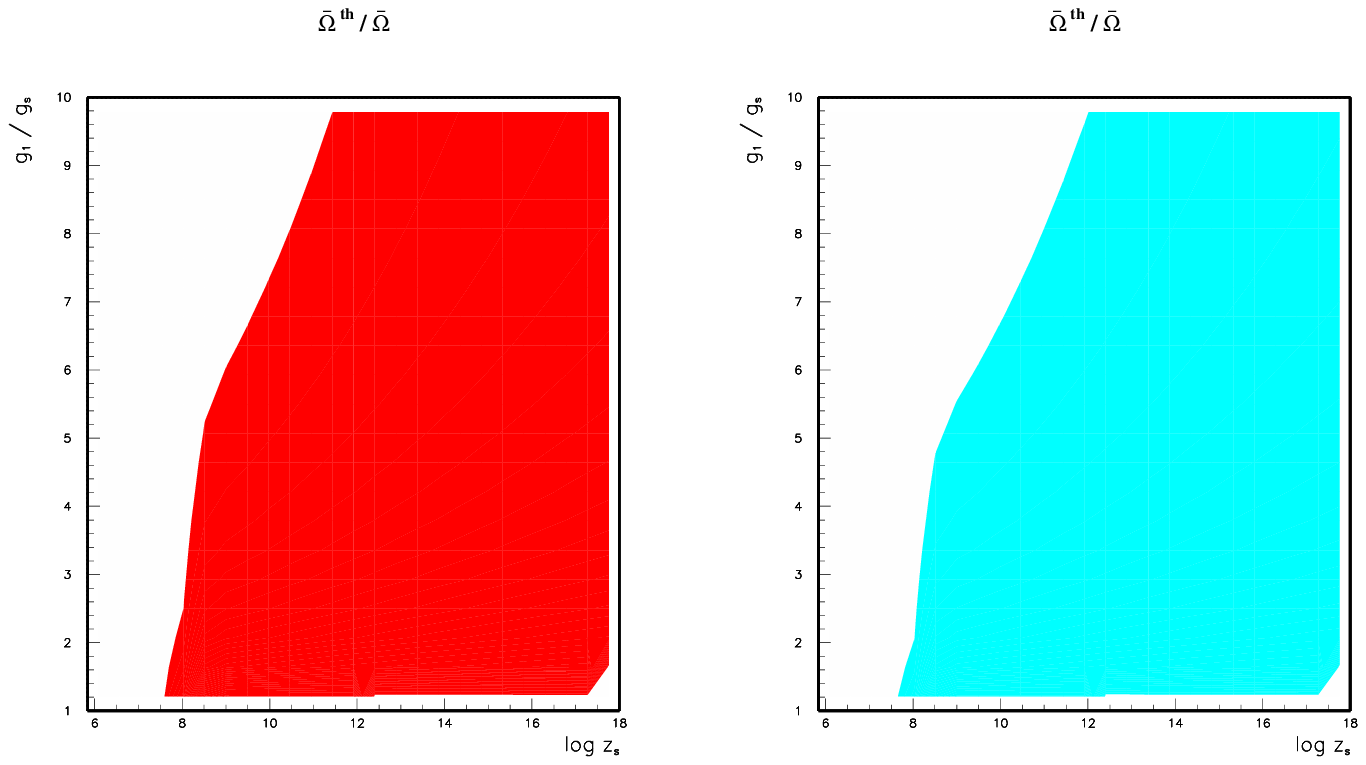}}
\vspace*{-18.0cm}
\caption[a]{The visibility region of the VIRGO pair in 
the case of  a noise reduction vector $\vec{\rho}=(0.1,0.01,1)$. At the left 
the overlap is maximal whereas at the right the overlap is minimal. As 
in the previous plots $g_1 = 1/20$ and $n_r = 10^3$.}
\label{strrid}
\end{figure}
In Fig. \ref{ger} the meaning of the shaded regions is exactly the same 
as in Fig. \ref{ita} but with the assumption of minimal overlap. In fact, 
the relevant profile of $\gamma(f)$ is the one labeled with C in 
Fig. \ref{over}. By comparing Fig. \ref{ita} with Fig. \ref{ger} we can 
notice that the effects of the maximization of the overlap is to increase 
the visibility region both in terms of $\bomax/\bom$ and in terms of 
$\bomth/\bom$. 
For sake of completeness we also report in Fig. \ref{fra} the patterns 
of the visibility region for non-scale-invariant spectra in the case of 
intermediate overlap. In this case the two VIRGO detectors are assumed 
to be roughly 500 km apart and this corresponds to the profile B of 
Fig. \ref{over}. As we can see from Fig. \ref{fra} the area of the 
visibility region is, approximately, in between the ones obtained in 
the case of Figs. \ref{ita} and \ref{ger}. 

In order to complete our analysis we would like to discuss the 
simultaneous effect of overlap maximization and noise reduction since 
this is one of the open possibilities of the upgraded VIRGO program. 
Suppose, for instance, that the noise configuration of the upgraded 
VIRGO would correspond to a reduction vector 
$\vec{\rho}= (0.1, 0.01, 1)$. If the overlap is either maximal or 
minimal the visibility region are the ones reported in 
Fig. \ref{strrid}. In particular, in Fig. \ref{strrid} we report the 
regions for which $\bomth > \bom$ (needless to say that we report 
only those regions of parameter space for which the BBN bound 
is satisfied, i.e. $\bomth < \bomax$). 
In the left plot of Fig. \ref{strrid} we assumed 
 maximal overlap (profile A of Fig. \ref{over}) whereas in the 
right plot we assumed minimal overlap (profile C of Fig. \ref{over}).
As we can see, a joined reduction of thermal noises overwhelms almost 
completely the effect of the maximization of the overlap in the sense
that, if the thermal noises are consistently reduced, the visibility
region seems to be rather insensitive to the relative location 
of the two detectors. Similar conclusions (and similar plots) 
are obtained in the case of different reduction vectors 
affecting the thermal noises. Finally, if the shot noise 
contribution is reduced the sensitivity is only mildly 
affected \cite{us}. The reason for this statement stems from the 
fact that the shot noise reduction starts being  important for 
frequencies larger than the kHz (see Fig. \ref{noise}) \cite{us}.
But for $f > 1$ kHz the overlap reduction is very efficient in spite 
of the location of the two detectors as it can be argued from Fig. 
\ref{over}.

 We close this section with some considerations about the 
magnitude of the ratios $\bomax/\bom$ and $\bomth/\bom$. 
Let us  start from the case without noise reduction 
($\vec{\rho}\,=\,(1,1,1)$). In the case of correlation A, the 
ratios $\bomax/\bom$ and $\bomth/\bom$ get their maximal values, 
of the order of 3 and 2, respectively, in the region 
$g_1/g_s\,\sim\,1$ and $\log{z_s}\,\sim\,9$. By inserting  
the fiducial values $g_1 = 1/20$ and $n_r = 10^3$ in Eq. (\ref{theo}) 
one obtains $h_0^2\,\bomth\,\simeq\,1.7\,\times\,10^{-7}$ and, thus, 
the minimum detectable $h_0^2\,\bom$ turns out of the order of 
$8.5\,\times \,10^{-8}$. This sensitivity gets worse if we move the 
second detector  from A 
to C. The loss in sensitivity is 
roughly of the same order as in the case of the flat 
spectrum (see Tab. 1).

The reduction of noise does not affect the region of the 
($g_1/g_s,\,\log{z_s}$)-plane where the ratios $\bomax/\bom$ and 
$\bomth/\bom$ are maximal. On the contrary, the 
magnitude of these ratios is affected: for the correlation A and in 
the case $\vec{\rho}\,=\,(0.1,0.01,1)$, 
$\bomax/\bom$ and $\bomth/\bom$ can be, respectively, 
 as large as 40 or 30. This implies that the minimum detectable $h_0^2\,\bom$ 
is of the order of $3.4\,\times\,10^{-9}$, roughly 25 times less than 
the corresponding value without noise reduction. This improvement 
in sensitivity is about twice
the one achieved in the same 
comparison but for the case of flat spectrun (see Tab. 1). The same 
considerations hold if we repeat this analysis to the case of the 
correlation C.

\section{Conclusion}

The effect of maximization (or reduction) of the overlap of two coaligned 
VIRGO detectors depends upon the specific form of the logarithmic energy 
spectrum of relic GW backgrounds. If the spectra are 
purely scale-invariant the effect of minimization of the overlap are 
negligible. If the spectra are non-scale-invariant the maximization of 
the overlap has a sizable impact on the sensitivity. 

If we simultaneously maximize the overlap and reduce (selectively) the 
contributions of the pendulum and pendulum's internal modes to the noise 
power spectra, the sensitivity of the VIRGO pair to a scale-invariant 
spectrum can be as low as $10^{-9}$ after one year of observations and 
with SNR = 1. 

In the assumption that no noise reduction will take 
place in the context of the upgraded VIRGO 
program, the maximization of the overlap alone  
 enlarges sizably the visibility window
(always for $T = 1$ yr and SNR = 1)  
for the case of non-scale-invariant (and growing with frequency) 
logarithmic energy spectra.
In the assumption of a consistent reduction of the thermal noises
the sensitivity to growing logarithmic energy spectra increase 
significantly in spite of the location of the second VIRGO 
interferometer since, in the latter case, the good effect of the 
overlap maximization is overcome by the thermal noise reduction. 

From a theoretical perspective, our results support the conclusion
that the maximization of the overlap has a different impact depending 
upon the analytical form of the logarithmic energy spectrum, urging, in 
particular dedicated studies of the forthcoming data for 
the specific case of growing logarithmic energy spectra. From a purely 
experimental perspective our findings suggest that a simultaneous 
maximization of the overlap and a noise reduction in the pendulum and 
pendulum's internal modes is desirable and extremely useful for a 
decisive improvement in the sensitivity of the VIRGO pair. 

\section*{Acknowledgments}

We want to thank A. Giazotto for very interesting remarks and 
conversations.

\end{document}